\documentclass[aps,prd,amsmath,amssymb,nofootinbib,notitlepage]{revtex4-1}
\usepackage{color}
\usepackage{graphicx}
\usepackage{bm}
\usepackage{bbm}

\usepackage{amsfonts}

\usepackage{pst-grad}
\usepackage{dsfont}
\usepackage{tikz}
\usetikzlibrary{shapes,arrows,backgrounds}
\usepackage{placeins}

\newcommand{\erw}[1]{\left \langle #1 \right \rangle}

\newcommand{\komm}[2]{\ensuremath{\left[\,#1,#2\,\right]}}

\newcommand{\algebra}[1]{\mathfrak{#1}}
\newcommand{\group}[1]{\mathcal{#1}}
\newcommand{\gU}{\mathcal{U}}

\newcommand{\cC}{\mathcal C}
\newcommand{\Z}{\mathbb{Z}}
\newcommand{\id}{\mathbbm{1}}

\newcommand{\vev}[1]{\left\langle #1 \right\rangle}

\newcommand{\rep}{\mathcal{R}}

\newcommand{\trP}{\mathcal{P}}

\newcommand{\trnsp}{\mathsf{T}}

\DeclareMathOperator{\tr}{tr}

\graphicspath{{.}}
\begin{document}
\title{Casimir Scaling and String Breaking in $\bm{G_2}$ Gluodynamics}

\author{Bj\"orn H. Wellegehausen}
\author{Andreas Wipf}
\author{Christian Wozar}
\thanks{\texttt{Bjoern.Wellegehausen@uni-jena.de},
\texttt{A.Wipf@tpi.uni-jena.de},
and \texttt{Christian.Wozar@uni-jena.de}}
\affiliation{Theoretisch-Physikalisches Institut,
Friedrich-Schiller-Universit{\"a}t Jena, Max-Wien-Platz 1, 07743
Jena, Germany}

\begin{abstract}
\noindent
We study the potential energy between static charges in
$G_2$ gluodynamics in three and four dimensions. Our work
is based on an efficient local hybrid Monte-Carlo algorithm
and a multi-level L\"uscher-Weisz algorithm with exponential
error reduction to accurately measure expectation values
of Wilson- and Polyakov loops. Both in three and four
dimensions we show that at intermediate scales the string 
tensions for charges in various $G_2$-representations scale with 
the second order Casimir. In three dimensions Casimir scaling is
confirmed within one percent for charges in representations
of dimensions $7,14,27,64,77,77',182$ and $189$
and in $4$ dimensions within $5$ percent
for charges in representions of dimensions $7,14,27$ and $64$.
In three dimensions we detect string breaking for charges in the 
two fundamental representations. The scale for string
breaking agrees very well with the mass of the created pair 
of glue-lumps.
\end{abstract}

\maketitle

\section{Introduction}

\noindent
There is compelling experimental evidence that the fundamental
constituents of QCD, quarks and gluons, never show up as asymptotic
states of strong interaction -- rather they are confined in mesons and
baryons. Understanding the dynamics of this confinement mechanism 
is one of the challenging problems in strongly coupled gauge theories.
There are convincing analytical and numerical
arguments to believe that  confinement is a property of pure gauge theories 
(gluodynamics) alone and that the underlying mechanism should not depend on 
the number $N$ of colours. Confinement is lost at high temperatures
and for gauge groups with a non-trivial center the trace of the Polyakov loop 
\begin{equation}
P(\vec{x})=\tr \trP(\vec{x}),\quad \trP(\vec{x})=\frac{1}{N}\tr 
\left(\exp\; i\int_0^{\beta_T}
A_0(\tau,\vec{x}) \,d\tau\right),\quad \beta_T=\frac{1}{T},\label{intro1}
\end{equation}
vanishes in the confined low-temperature phase and is close to an element of
the center in the deconfined high-temperature phase. In gluodynamics 
or gauge theories with matter in the adjoint representation
the action and measure are both invariant under 
\emph{center transformations}, whereas the Polyakov loop 
transforms non-trivially and hence serves as order parameter for 
the global center symmetry. This means that the center symmetry is 
realized in the confined phase and spontaneously broken in the 
deconfined phase.

In the vicinity of the transition point the dynamics of the Polyakov loop
is successfully described by effective $3d$ 
scalar field models for 
the characters of the Polyakov loop
\cite{Svetitsky:1982gs,Yaffe:1982qf,Wozar:2006fi,Wellegehausen:2009rq}. If one
further projects the scalar fields onto the center of the gauge group then one arrives at generalized Potts models describing the effective
Polyakov-loop dynamics \cite{wipf:2006wj}. The temperature dependent
couplings constants of these effective theories have been calculated ab initio 
by inverse Monte Carlo methods in \cite{Wozar:2006fi}.

With dynamical quarks in  the fundamental representation the center 
symmetry is \emph{explicitly broken} and the Polyakov loop
points always in the direction of a particular center element. In a strict sense 
the Polyakov loop ceases to be an order parameter. This is attributed to 
breaking of the string connecting  a static `quark anti-quark pair' when one 
tries to separate the charges. It breaks via the spontaneous creation of 
dynamical quark anti-quark pairs which in turn screen the individual static 
charges.

The pivotal role of the center for confinement also follows
from a recent observation relating the Polyakov loop with
center averaged spectral sums of the Dirac operator 
\cite{Gattringer:2006ci,Synatschke:2007bz,Bilgici:2008qy}. More
precisely, for gauge groups with \emph{non-trivial center} one can relate 
the expectation value of the Polyakov loop to dual condensates.
This result could finally explain why for gauge groups with a non-trivial center and 
fundamental matter the transition temperatures for the 
deconfinement and chiral phase transitions coincide. On the contrary, for gauge 
theories with adjoint matter the two transition temperatures
can be very different \cite{Kogut:1985xa,Karsch:1998qj}.

To clarify the relevance of the center for confinement it suggests itself 
to study \emph{pure gauge theories} whose gauge groups have a trivial center. 
For such theories the string connecting external charges can break 
via the spontaneous creation of dynamical `gluons'  such that the Polyakov 
loop acquires a non-vanishing expectation value for all temperatures,
similarly as it does in QCD with dynamical fermions. Here the simple 
gauge group $SO(3)$ suggests itself and indeed the $SO(3)$ gauge theory 
has been studied in great detail on the lattice, see for example \cite{wipf:2002sk}. 
Unfortunately, via the non-trivial first homotopy group $\pi_1(SO(3))=\Z_2$ 
the lattice gauge theory `detects' its simply connected universal covering 
group $SU(2)$. To avoid the resulting lattice artifacts one should investigate 
theories with simply connected gauge groups with trivial center.

\begin{table}\label{tabelcenter}
\caption{\label{tab:centersGenerators}Centers ${\cal Z}$ of simple lie groups.}
\begin{ruledtabular}
\begin{tabular}{lcccccccccc}
group& $A_r$  &$B_r$   &$C_r$  &$D_r,\;r\hbox{ even}$  & $D_r,\;r\hbox{ odd} 
$ &  $E_6$& $E_7$& $E_8$& $F_4$&$G_2$
 \\ \hline
center $\mathcal{Z}$ & $\Z_{r+1}$ & $\Z_2$ & $\Z_2$ & $\Z_2\times \Z_2$ & $\Z_4
$ & $\Z_3$&$\Z_2$&$\id$&$\id$&$\id$\\
\end{tabular}
\end{ruledtabular}
\end{table}
From Tab.~\ref{tab:centersGenerators}, taken from \cite{Ford:1998sy}, one reads
off that the smallest simple Lie group with these properties is the
$14$-dimensional exceptional Lie group $G_2$.
This is one reason why the group in Bern  investigated
$G_2$ gauge theories with and without Higgs fields in series of
papers \cite {Holland:2003kg, Holland:2003jy,Pepe:2006er}. In their
pioneering works it has been convincingly demonstrated that
$G_2$-gluodynamics shows a first order finite temperature phase
transition without order parameter from a confining to a deconfining phase.
In this context confinement refers to confinement at intermediate scales, where 
a Casimir scaling of string tensions has been reported \cite{Liptak:2008gx}. 
On large scales strings will finally break due to spontaneous gluon production 
and the static inter-quark potential is expected to flatten \cite{Greensite:2003bk}.
However, the threshold energy for string breaking in $G_2$-gauge theory 
is rather high and all previous attempts to detect this flattening have been 
without success. In the present paper we shall demonstrate that string breaking 
for charges in the fundamental and adjoint representations of $G_2$ takes place 
at the expected scales. To that aim we implemented a
slightly modified L\"uscher-Weisz multistep algorithm
for high-precision measurements of the static inter-quark potential.

The present paper deals with $G_2$-gluodynamics in $3$ and $4$
dimensions. The simulations are performed with an efficient
and fast implementation of a local HMC algorithm.
Below we shall calculate the potentials at intermediates scales 
for static charges in the $7, 14,27,64,77,77',182$ and $189$-dimensional 
representations. We show that in $3$ and $4$ dimensions the
string tensions on intermediate scales are proportional to the second 
order Casimir of the  representations. The high-precision measurements in 
$3$ dimensions confirm Casimir scaling within $1$ percent. 
In $4$ dimensions Casimir scaling for the lowest $4$ representations
is fulfilled within $5$ percent.
In $3$ dimensions we also calculated the static potential for widely
separated charges  in the two fundamental representations. 
In both cases we see a flattening of the potential which signals the 
breaking of the connecting string. The energy
where string breaking sets in is in full agreement with the 
independently calculated masses of the glue lumps formed after 
string breaking.

%============================================================================

\section{The group $\bm{G_2}$}
\label{sect:g2}
\noindent
The exceptional Lie-Group $G_2$ is the automorphism group of the octonion
algebra or, equivalently, the subgroup of $SO(7)$ that preserves 
any vector in its 8-dimensional real spinor representation. This
means that the $8$-dimensional real spinor representation of Spin$(7)$ branches
into the trivial representation and the $7$-dimensional fundamental 
representation of $G_2$.  The $14$-dimensional fundamental representation
of $G_2$, which at the same time is the adjoint representation, 
arises in the branching of the adjoint of $SO(7)$ according to $21\to 7\oplus 14$.
The $27$-dimensional representations of $SO(7)$ acting on
symmetric traceless  $2$-tensors remains irreducible under $G_2$. 
In this  work we need the following \emph{branchings} of $SO(7)$-representations 
to $G_2$:
\begin{equation}
7\rightarrow 7,\quad
21\rightarrow 14\oplus 7,\quad
27\rightarrow 27 ,\quad
35\rightarrow 27 \oplus 7 \oplus 1,\quad
77\rightarrow 77.
\end{equation}
For explicit calculations it is advantageous to view the elements of the 
$7\-$-dimensional representation of $G_2$ as matrices in the defining 
representation of $SO(7)$, subject to seven independent cubic 
constraints \cite{Holland:2003jy}:
\begin{equation}
\label{eq:g2constraint}
T_{abc} = T_{def}\,g_{da}\,g_{eb}\,g_{fc}.
\end{equation}
Here $T$ is a total antisymmetric tensor given by
\begin{equation}
T_{127} = T_{154} = T_{163} = T_{235} = T_{264} = T_{374} = T_{576} = 1.
\end{equation}
%The cubic constraints defining the elements of the group 
%reduce the $21$ generators of $SO(7)$ to $14$ generators of $G_2$. 
%The Weyl group of $G_2$ is the dihedral group $D_6$ of order $12$.
The gauge group $SU(3)$  of strong interaction is a subgroup of $G_2$ 
and the corresponding coset space is a sphere \cite{Macfarlane:2002hr},
\begin{equation}
G_2/SU(3) \sim S^6.
\end{equation}
This means that every element $\gU$ of $G_2$ can be factorized as
\begin{equation}
\gU=\group{S} \cdot \group{V} \quad \text{with} \quad \group{V} \in SU(3) \quad \text{and} \quad \group{S} \in G_2/SU(3),\label{decomposition}
\end{equation}
and we shall use this decomposition in our simulations.
The short exact sequence
\begin{equation}
0=\pi_4(S^6)\to \pi_3(SU(3))\to \pi_3(G_2)\to \pi_3(S^6)=0
\end{equation}
shows that $\pi_3(G_2)=\Z$ and hence there should exist 
$G_2$-instantons of any integer topological charge. In the 
charge $k$-sector 
there are at least $3k$ magnetically charged defects \cite{Ford:1998sy}. 

%=================================
Any irreducible representation of $G_2$ is characterized by its highest 
weight vector $\mu$ which is a linear combination of the fundamental weights,
$\mu=p\mu_{(1)}+q\mu_{(2)}$, with non-negative integer coefficients $p,q$ called
Dynkin labels.  The dimension of an arbitrary irreducible representation 
$\rep=[p,q]$ can be calculated with the help of Weyl's dimension formula and is given by
\begin{equation}
d_\rep\equiv\hbox{dim}_{p,q}=\frac{1}{120}(1+p)(1+q)(2+p+q)(3+p+2q)(4+p+3q)(5+2p+3q).
\end{equation}
Below we also use the physics-convention and denote 
a representation by its dimension. For example, the fundamental representations are
$[1,0]=7$ and $[0,1]=14$. However, this notation is
ambiguous, since there exist different representations with the
same dimension. For example $[3,0]=77$ and $[0,2]=77'$
have the same dimension. An irreducible representation of $G_2$ can
also be characterized by the values of the two Casimir operators
of degree $2$ and $6$. Below we shall need the values of the quadratic 
Casimir in a representation $[p,q]$, given by
\begin{equation}
\cC_\rep\equiv\cC_{p,q}=2p^2+6q^2+6pq+10p+18q.
\end{equation}
For an easy comparison we normalize these `raw' Casimir values with respect to
the defining representation by $\cC'_{p,q}=\cC_{p,q}/\cC_{1,0}$. The
normalized Casimir values for the eight non-trivial representations with smallest dimensions 
are given in Tab.~\ref{tab:representationCasimirs}.
\begin{table}
\caption{\label{tab:representationCasimirs} Representations of $G_2$ with
corresponding dimension and Casimir values.}
\begin{ruledtabular}
\begin{tabular}{lcccccccc}
representation $\rep$ & $[1,0]$ & $[0,1]$ & $[2,0]$ & $[1,1]$ & $[3,0]$ &
$[0,2]$ & $[4,0]$ & $[2,1]$ \\ \hline
dimension $d_\rep$ & $7$ & $14$ & $27$ & $64$ & $77$ & $77'$ & $182$ & $189$ \\ 
Casimir eigenvalue $\cC_\rep$ & $12$ & $24$ & $28$ & $42$ & $48$ & $60$ & $72$ &
$64$ \\
Casimir ratio $\cC'_\rep$ & $1$ & $2$ & $7/3$ & $3.5$ & $4$ & $5$ & $6$ &
$16/3$\\
\end{tabular}
\end{ruledtabular}
\end{table}

Quarks and gluons in $G_2$ are in the fundamental representions
 $7$ and $14$, respectively.
To better understand $G_2$-gluodynamics we recall the decomposition of
tensor products of these representations,
\begin{equation}
\label{eq:representationsG2}
\begin{aligned}
7 \otimes 7&=1 \oplus 7 \oplus 14 \oplus27\\ 
7 \otimes14&=7 \oplus27 \oplus 64\\ 
14 \otimes14&=1 \oplus14 \oplus 27 \oplus 77 \oplus 77' \\
7 \otimes 7 \otimes 7&=1 \oplus 4 \cdot 7 \oplus 2 \cdot14 \oplus 3 \cdot27 
\oplus 2 \cdot 64 \oplus 77'\\
14 \otimes14\otimes14&=1\oplus 7 \oplus 5\cdot14\oplus 
3\cdot27\oplus\dotsb
\end{aligned}
\end{equation}
The decompositions \eqref{eq:representationsG2} show that, similarly as in QCD,
two or three quarks or two or three gluons can build
colour singlets -- mesons, baryons or glueballs. 
Since three gluons can screen the charge of a single (static) quark,
\begin{equation}
7 \otimes14 \otimes14 \otimes14=1 \oplus \dotsb\,,\label{eq:representationsG2p}
\end{equation}
one expects that the string between two static quarks 
will break for large charge separations. The two remnants are 
two glue-lumps -- charges screened by  (at least) $3$ gluons. The same 
happens for charges in the adjoint representation. Each adjoint charge can be screened by
one gluon.

\subsection*{Construction of characters from tensor products}
\noindent
The character $\chi_\rep=\tr\rep$ of any irreducible representation $\rep$ 
is a polynomial of the characters $\chi_7$ and $\chi_{14}$ of the two fundamental representations
$7$ and $14$. For example, the first two decompositions in
\eqref{eq:representationsG2} imply
\begin{equation}
\begin{aligned}
\chi_{27}&=\chi_7\cdot\chi_7-\chi_1-\chi_7-\chi_{14}\\
\chi_{64}&=\chi_7\cdot\chi_{14}-\chi_{7}-\chi_{27}=\chi_7\chi_{14}-\chi_7^2+\chi_1+\chi_{14}
\end{aligned}
\end{equation}
and yield the characters of the representations $27$ and $64$
as polynomials of $\chi_7$ and $\chi_{14}$. From further tensor products
of irreducible representions one can calculate the polynomial in 
$\chi_\rep=$Pol$_{\rep}(\chi_7,\chi_{14})$ for any irreducible representation $\rep$.
For a fast implementation of our algorithms we also need reducible representations.
In particular we use
\begin{equation}
\quad (7\otimes7)_\text{s}, \quad (7\otimes 7 \otimes 7)_\text{s},\quad
(7\otimes7\otimes7\otimes7)_\text{s},\quad
\left(7\otimes7\right)_\text{s}\otimes14
\end{equation}
where the subscript `$\text{s}$' denotes the symmetrized part of the respective
tensor product. Comparing the reduction of representations for $SO(7)$ and $G_2$ and
mapping representations from $SO(7)$ to $G_2$ the following characters of
reducible representations can be computed
\begin{equation}
\begin{aligned}
\chi_{(7\otimes 7)_\text{s}}&=\chi_{27}+\chi_{1},\\
\chi_{\left(7\otimes7\otimes7\right)_\text{s}}&=\chi_{77}+\chi_{7},
\\
\chi_{\left(7\otimes7\otimes7\otimes7\right)_\text{s}}
&=\chi_{182}
+\chi_{77}+\chi_{27}+\chi_{64}+2\,\chi_{14}+\chi_{7},\\
\chi_{\left(7\otimes7\right)_\text{s}\otimes14}&=\chi_{189}+\chi_{27}+\chi_{1}.
\end{aligned}
\end{equation}

\section{Casimir scaling and string breaking for $\bm{SU(N)}$ gauge theories}
\noindent
In QCD quarks and anti-quarks can only be screened by particles with 
non-vanishing $3$-ality, especially not by gluons. Thus, in 
zero-temperature \emph{gluodynamics} the potential energy for two static 
color charges is linearly rising up to arbitrary large separations of the
charges. The potentials for charges in a representation
$\rep$ can be extracted from the 2-point correlator of 
Polyakov loops or the expectation values of Wilson loops with 
time-extent $T$ according to
\begin{equation}
\langle P_\rep(0)P_\rep(R)\rangle=e^{-\beta_T V_\rep(R)}\quad,\quad
\langle W_\rep(R,T)\rangle=e^{\kappa_R-TV_\rep(R)}.\label{potposs}
\end{equation}
With dynamical quarks the string should break at a characteristic 
length $r_b$ due to the spontaneous creation of quark anti-quark pairs
from the energy stored  in the flux tube connecting the static charges. 
However, for intermediate  separations $r<r_b$ the string cannot break 
since there is not enough energy stored in the flux tube. 

For \emph{pure gauge theories} we expect the 
following qualitative behavior of the static potential: 
At short distances perturbation theory applies and the interaction is 
dominated by gluon exchange giving rise to a Coulomb-like potential, $V\sim -\alpha/r$,
the strength $\alpha$ being proportional to the value 
$\cC_\rep$ of the quadratic Casimir 
operator in the given representation $\rep$ of the charges;
at intermediate distances, from the onset of confinement to the onset
of color screening at $r_b$, the potential is expected to be linearly rising,
$V\sim\sigma r$, and the corresponding string tension is again proportional to the quadratic
Casimir; at asymptotic distance scales (partial) screening sets in such
that the string tension typically decreases and only depends on the $N$-ality 
of the representation. In particular for center-blind color
charges or gauge groups without center the potential flattens. 
The characteristic length  $r_b$ where the intermediate confinement regime turns 
into the asymptotic screening regime is determined by the masses of the debris 
left after string breaking. The Casimir scaling hypothesis, according to
which  the string tension at intermediate scales is proportional to the 
quadratic Casimir of the representation \cite{Oleson:1984}, is exact for
two dimensional continuum and lattice gauge theories and
dimensional reduction arguments support that it also holds in higher 
dimensions. Within the Hamiltonian approach to Yang-Mills theories
in $2+1$ dimensions the following prediction for the string tensions
has been derived \cite{Karabali:2009rg}
\begin{equation}
\sigma_\rep=\frac{g^4}{4\pi}\cC_{14}\cC_{\rep}.\label{Nair}
\end{equation}
For \emph{pure SU(2) and SU(3) gauge theories} in three and four dimensions there 
is now conclusive numerical evidence for \emph{Casimir scaling} from Monte-Carlo simulations:
for $SU(2)$ in $3$ dimensions
\cite{Oleson:1984,Poulis:1995}
and in $4$ dimensions \cite{Ambjorn:1984,Michael:1985,Griffiths:1985,Trottier:1995}
as well as for $SU(3)$ in $4$ dimensions at finite temperature
\cite{Muller:1992} 
and zero temperature \cite{Campbell:1986,
Michael:1992,Bali:2000,Piccioni:2005}.
In particular the simulations for $SU(3)$ gluodynamics 
in \cite{Bali:2000} confirm Casimir scaling within $5\%$ for 
separations up to $1$ fm of static charges in representations with 
Casimirs (normalized by the Casimir of $\{3\}$) up to $7$. \emph{String breaking}
for charges in the adjoint representation has been found in several simulations:
In $3$-dimensional $SU(2)$-gluodynamics with improved action 
and different operators in \cite{Philipsen:1999,Stephenson:1999} and in
$4$-dimensional $SU(2)$-gluodynamics in \cite{Forcrand:2000} with
the help of a variational approach involving string and glueball operators. 
For a critical discussion of the various approaches we refer to \cite{Kratochvila:2003},
where string breaking in a simple setting but with an improved
version of the L\"uscher-Weisz algorithm has been analyzed and compared
with less sophisticated approaches. 
There is a number of works in which a violation of  Casimir scaling 
on intermediate scales have been reported. For example, it has been
claimed that in $4$-dimensional $SU(N)$-gluodynamics with larger $N=4,6$ 
the numerical data favor the sin-formula, as suggested by supersymmetry,
in place of the Casimir scaling formula \cite{DelDebbio:2002}. 
The differences between the Casimir scaling law  and sin-formula are tiny and it 
is very difficult to discriminate between the two predictions in numerical
simulations. Indeed, in \cite{Lucini:2001} agreement with Casimir scaling and
sin-formula in $4$-dimensions and disagreement in $3$-dimensions has been claimed. 
In addition the high precision simulation based on the L\"uscher-Weisz algorithm 
in \cite{Kratochvila:2003b} point to a violation of the Casimir scaling law in 
$3$-dimensional $SU(2)$ gluodynamics. 
In a very recent paper Pepe and Wiese \cite{Pepe:2009} reanalyzed
the static potential for $SU(2)$-gluodynamics in $3$ dimensions with
the help of the L\"uscher-Weisz algorithm and confirmed Casimir scaling
at intermediate scales and $2$-ality scaling at asymptotic scales.

For \emph{gauge theories with matter} we expect a similar qualitative 
behavior: a Coulomb-like potential at short distances,
Casimir scaling at intermediate distances and (partial) screening at asymptotic
distances. The string tension at asymptotic scales depends both
on the $N$-alities of the static color charges and of
the dynamical matter. In particular, if dynamical quarks or
scalars can form center blind composites
with the static charges then the potential is expected to flatten 
at large separations.  
To see any kind of screening between fundamental charges
requires a full QCD simulation with sea quarks, which is demanding.
Thus the earlier works dealt with gauge theories with scalars
in the fundamental representation. For example, in \cite{Philipsen:1999a}
clear numerical evidence for string breaking in the $3$-dimensional 
$SU(2)$-Yang-Mills-Higgs model via a mixing analysis of string and two-meson operators 
has been presented. Probably the first observation of hadronic string breaking in
simulation of QCD$_3$ with two flavors of dynamical staggered
fermions using only Wilson loops have been reported in
\cite{Trottier:1999,Trottier:2005}. Despite extensive searches for 
colour screening in $4$-dimensional gauge theories with dynamical fermions 
the results are still preliminary at best. First indications for string breaking in 
two-flavor QCD, albeit only at temperatures close to or above the 
critical deconfinement temperature, have been reported in \cite{TeTar:1998}. 
More recently Bali et al.\ used sophisticated methods
(e.g.\ optimized smearing, improved action, stochastic estimator
techniques, hopping parameter acceleration) to resolve string breaking in
$2$-flavor QCD at a value of the lattice spacing $a^{-1}\approx 2.37$~GeV
and of the sea quark mass slightly below $m_s$ \cite{Bali:2005}. By extrapolation they estimate
that in real QCD with light quarks the string breaking should happen at
$r_b\approx 1.13$~fm.

To measure the static potential and study string breaking three approaches have 
been used: correlations of Polyakov loops at finite temperature,
variational ansaetze using two types of operators (for the string-like states and
for the broken string state) and Wilson loops. Most results on Casimir
scaling and string breaking have been obtained with the first two methods.
This is attributed to the small overlap of the Wilson loops with the broken-string 
state. To measure Polyakov or Wilson loop correlators for 
charges in higher representations or to see screening at asymptotic
scales one is dealing with extremely small signals down to $10^{-40}$.
In order to measure such small signals one needs to improve existing
algorithms considerably or/and use improved versions of 
the L\"uscher-Weisz multistep algorithm.

\begin{figure}[htb]
\input{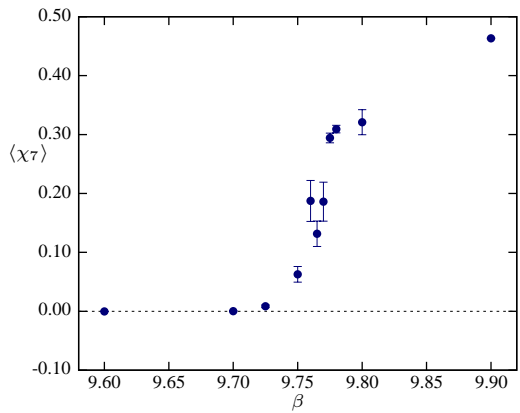}
\caption{Phase transition on a $16^3 \times 6$ lattice in terms of the Polyakov
loop in the fundamental representation.}
\label{fig:ymTransition}
\end{figure}

For \emph{gauge groups with trivial centers} like $G_2,\, F_4$ or $E_8$ the flux 
tube between static charges in any representation will always break 
due to gluon production. The potential flattens
for large separations and expectation values of the Polyakov loop 
never vanish \cite{Pepe:2006er}.
However, for $G_2$ it changes rapidly at the phase transition temperature 
and is very small  in the low-temperature confining phase, 
see Fig.~\ref{fig:ymTransition}. Similarly as in QCD we characterize 
confinement as the absence of free colour charges in the 
physical spectrum \cite{Greensite:2006sm,Liptak:2008gx}.

\FloatBarrier

\section{Algorithmic considerations}
%----------------------------------------------------------------------------
\subsection{Local hybrid Monte-Carlo}
\noindent
In simulations of gauge field theories different algorithms are in use. For 
$SU(N)$-gluodynamics heat-bath algorithms based on the Cabibbo-Marinari $SU(2)$ 
subgroup updates, often improved by over-relaxation steps,
have proven to be fast and reliable. For QCD with dynamical fermions 
a hybrid Monte-Carlo (HMC) scheme is preferable. Based on
\cite{Marenzoni:1993im} also \emph{local} versions of HMC algorithms 
are available where single links are evolved in a HMC style. 
According to \cite{Gehrmann:1999wr} the cost for the local hybrid 
Monte-Carlo (LHMC) is about three times more than for a combined 
heat-bath and overrelaxation (HOR) scheme for the case of $SU(N)$-gluodynamics.

For the exceptional gauge group $G_2$ there exists a modification of the
heat-bath update \cite{Pepe:2006er} which combines the heat-bath update
for a $SU(3)$-subgroup with randomly distributed $G_2$ gauge transformations 
to rotate the $SU(3)$ subgroup through $G_2$. In the present work we instead
use a LHMC algorithm for several good reasons: 
First, the formulation is given entirely in terms of Lie-group and 
Lie-algebra elements and there is no need to back-project onto $G_2$.
The autocorrelation time can be controlled (in certain ranges) by the integration 
time in  the molecular dynamics part of the HMC algorithm.
Furthermore, one can use a real representation of $G_2$ and 
relatively simple analytical expressions for the two involved  exponential maps 
to obtain a  fast implementation of the algorithm.
Finally, the inclusion of a (normalized) Higgs field is straightforward and  does 
not suffer from a  low Metropolis acceptance rate (even for large hopping parameters).

The LHMC algorithm has been essential for obtaining the results
in the present work. Since we developed the first implementation for $G_2$
it is useful to explain the technical details for this
exceptional group. As any (L)HMC algorithm for gauge theories it is based on a fictitious dynamics 
for the link-variables on the gauge group manifold. The ``free evolution'' on a 
semisimple group is the Riemannian geodesic motion with respect to the Cartan-Killing metric
\begin{equation}
ds^2_G=\kappa\tr\left(d\gU \gU^{-1}\otimes d\gU\gU^{-1}\right).
\end{equation}
In the fictitious dynamics the interaction term is given by the Yang-Mills action
of the underlying lattice gauge theory and hence it suggests itself to derive
the dynamics from the Lagrangian
\begin{equation}
L=\frac{1}{2}  \sum \limits_{x,\mu}\tr
\left(i\,\dot{\gU}_{x,\mu}\gU_{x,\mu}^{-1}\right)^2-S_\text{YM}[\gU],
\end{equation}
where `dot' denotes the derivative with respect to the fictitious time 
parameter $\tau$ and
\begin{equation}
S_\text{YM}[\gU]=\frac{\beta}{2 \, N_c} \sum
\limits_{x,\mu\nu}\tr\left(2\,N_c-\gU_{x,\mu\nu}-\gU_{x,\mu\nu}^\dagger \right)
\end{equation}
is the Wilson action. The Lie algebra valued  fictitious \emph{conjugated 
link momentum} is given by
\begin{equation}
\algebra{P}_{x,\mu}=i\, \frac{\partial L}{\partial
\left(\dot{\gU}_{x,\mu}\gU_{x,\mu}^{-1}\right)}=i\,
\gU_{x,\mu}\frac{\partial L}{\partial \dot{\gU}_{x,\mu}}=-i\,
\dot{\gU}_{x,\mu}\gU_{x,\mu}^{-1}\,,
\end{equation}
and via a Legendre transform yields the pseudo-Hamiltonian
\begin{equation}
H=\frac{1}{2} \sum_{x,\mu}\tr \algebra{P}_{x,\mu}^2+S_\text{YM}[\gU].
\end{equation}
The equations of motion for the momenta are obtained by varying the
Hamiltonian. The variation of the Wilson action $S_\text{YM}[\gU]$
with respect to a fixed link variable $\gU_{x,\mu}$ is given by the corresponding
staple variable $R_{x,\mu}$, the sum of triple products of elementary link variables closing 
to a plaquette with the chosen link variable. Hence we obtain
\begin{equation}
\label{HMChamiltonian}
\begin{aligned}
\delta H &=
\sum \limits_{x,\mu} \tr
\left\{\algebra{P}_{x,\mu}\delta{\algebra{P}}_{x,\mu}-\frac{\beta}{2\,
N_c}\delta \gU_{x,\mu}\gU_{x,\mu}^\dagger
\left(\gU_{x,\mu}R_{x,\mu}-R_{x,\mu}^\dagger \gU_{x,\mu}\right)\right\} \\ 
&=\sum \limits_{x,\mu} \tr \algebra{P}_{x,\mu}\left\{\dot{\algebra{P}}_{x,\mu}
-F_{x,\mu}\right\}d\tau,
\qquad
F_{x,\mu}=\frac{i\,\beta}{2\, N_c}
\left(\gU_{x,\mu}R_{x,\mu}-R_{x,\mu}^\dagger \gU_{x,\mu}\right).
\end{aligned}
\end{equation}
The variational principle implies that the projection of the term between
curly brackets onto the Lie algebra $\algebra{g}_2$ vanishes,
\begin{equation}
\dot{\algebra{P}}_{x,\mu}=
F_{\mu,x}\big\vert_{\algebra{g}_2}.
\end{equation}
Choosing a trace-orthonormal basis $\{T_a\}$ of $\algebra{g}_2$
the equations for the (L)HMC dynamics can be written as follows,
\begin{equation}
\dot{\algebra{P}}_{x,\mu}=\sum \limits_a \tr \left(F_{x,\mu}T_a\right)T_a
\quad \text{and} \quad \dot{\gU}_{x,\mu}=i\,
\algebra{P}_{x,\mu}\gU_{x,\mu}
\end{equation}
with the ``force'' $F_{x,\mu}$ defined in \eqref{HMChamiltonian}.
Now a LHMC sweep consists of the following steps:
\begin{enumerate}
\item Gaussian draw of the momentum variable on a given link.
\item Integration of the equations of motion for the given link.
\item Metropolis accept/reject step,
\item Repeat these steps for all links of the lattice.
\end{enumerate}
This local version of the HMC does 
not suffer from an extensive $\delta H\propto V$ problem such that already
a second order symplectic (leap frog) integrator allows
for sufficiently large timesteps $\delta \tau$. In condensed form the
integration for a link variable yields
\begin{equation}
\gU(t+\delta\tau)=\exp\left(i \algebra{P}(t+\delta \tau/2) \delta
\tau\right)\gU(t).
\end{equation}
For a large range of Wilson couplings $\beta$ in our simulations an integration
length of $T=0.75$ with a step size of $\delta \tau=0.25$ is optimal for minimal
autocorrelation times and a small number of thermalisation sweeps. Acceptance
rates of more than $99\%$ are reached. Nevertheless, the most time consuming
part of the calculations involves the exponential maps. A calculation for $G_2$ can be
implemented fast and exact up to a given order in $\delta \tau$ as will be
shown in the next section.

\subsection{The exponential map $\bm{\algebra{g}_2\to G_2}$}\label{exmap}
\noindent
For an efficient and fast computation of the exponential map we use the
\emph{real embedding} of the $SU(3)$-representation $3\oplus\bar3$ into $G_2$, 
given by
\begin{equation}
\group{V}(\group{W})=\Omega^\dagger\begin{pmatrix}1&0&0\cr 0&\group{W}&0\cr 0&0&\group{W}^*\end{pmatrix}
\Omega\in G_2,\quad\hbox{with}\quad \group{W}\in SU(3).\label{embedding}
\end{equation}
One can choose the unitary matrix $\Omega$  to have block diagonal form
with $\Omega_{11}=1$. A possible choice for $\Omega$ is 
\begin{equation}
\Omega=\begin{pmatrix}1&0\cr 0&VQ\end{pmatrix}\quad\hbox{with}\quad
Q=\begin{pmatrix}
0&0&0&0&0&1\\ 0&0&1&0&0&0\\1&0&0&0&0&0\\0&0&0&0&1&0\\0&0&0&1&0&0\\0&1&0&0&0&0\\
  \end{pmatrix},\quad
V=\frac{1}{\sqrt{2}}\begin{pmatrix}
1 & i\\ i &1 \\   \end{pmatrix}\otimes \id_3.
\label{decompg2}
\end{equation}
Every element of $G_2$ can be factorized as
\begin{equation}
\gU=\group{S} \cdot \group{V}(\group{W}) \quad \text{with} 
\quad \group{S} \in G_2/SU(3).\label{groupdecomp}
\end{equation}
For a given timestep $\delta \tau$ in the molecular dynamics this factorization
will be expressed in terms of the Lie algebra elements with the help of  the exponential maps,
\begin{equation}
\exp \left \lbrace \delta \tau\, \algebra{u} \right \rbrace=\exp \left \lbrace
\delta \tau\, \algebra{s} \right \rbrace\cdot \exp \left \lbrace \delta \tau\,
\algebra{v} \right \rbrace \quad \text{with generators} \quad \algebra{u}\in\algebra{g}_2,\;\;
\algebra{v} \in \group{V}_*(\algebra{su}(3))\label{expalg}
\end{equation}
fulfilling the commutation relations
\begin{equation}
\komm{\algebra{v}}{\algebra{v^\prime}}=\algebra{v^{\prime \prime}} ,
\quad \komm{\algebra{v}}{\algebra{s}}=\algebra{s^{\prime}} \quad \text{and}
\quad
\komm{\algebra{s}}{\algebra{s^\prime}}=\algebra{v^\prime}+\algebra{s^{\prime \prime}}.
\end{equation}
The generators $\algebra{s}$ are orthogonal to the generators of the 
really embedded $SU(3)$-subgroup. To simplify the notation we absorb the time step $\delta \tau$ in
the Lie algebra elements.

The last exponential map in \eqref{expalg} can be calculated with the help
of the embedding \eqref{embedding} and the exponential map for
$SU(3)$, $\group{W}= \exp(\algebra{w})$,
which follows from the Cayley-Hamilton theorem for $SU(3)$-generators, 
see \cite{Laufer:1996fv}. The result can be
expressed in terms of the imaginary eigenvalues $w_1,w_2,w_3$ of $\algebra{w}$
and the differences $\delta_1=w_2-w_3,\,\delta_2=w_3-w_1$ and
$\delta_3=w_1-w_2$ as follows:
\begin{equation}
\group{W}=
\exp(\algebra{w})=-\frac{1}{\delta_1\delta_2\delta_3}
\left(\alpha_{\id}\id+\alpha_{\algebra{w}}\algebra{w}
+\alpha_{\algebra{w}^2}\algebra{w}^2
\right)
\end{equation}
with expansion coefficients
\begin{equation}
\alpha_\id=\sum_{i=1}^3 \delta_iw_{i+1}w_{i+2}e^{w_i},\quad
\alpha_{\algebra{w}}=\sum_{i=1}^3 \delta_iw_ie^{w_i},\quad
\alpha_{\algebra{w}^2}=\sum_{i=1}^3 \delta_i e^{w_i},
\end{equation}
wherein one identifies $w_{3+i}$ and $w_i$. 

For the generators $\{\algebra{u_1},\dots,\algebra{u}_{14}\}$ of $G_2$ we use the 
real representation given in  \cite{Greensite:2006sm}.  The
$\algebra{su}(3)$-subalgebra formed by the elements
$\{\algebra{u_1},\dots,\algebra{u}_{8}\}$ generates the really embedded $3\oplus\bar 3$ of $SU(3)$ and the remaining generators $\{\algebra{u_9},\dots,\algebra{u}_{14}\}$ 
generate  the coset-elements $\group{S}$ in the factorization
\eqref{groupdecomp}. With this choice for the generators the real
embedding \eqref{embedding} reads
\begin{equation}
\group{V}(\group{W})=\begin{pmatrix}1&0\cr 0&\group{V}_\perp\end{pmatrix},\qquad
\group{V}_\perp=
\begin{pmatrix}
a_{33}&-b_{33}& a_{32}& -b_{32}&-b_{31}&a_{31}\\
b_{33}&a_{33}& b_{32}& a_{32}&  a_{31}&b_{31}\\
a_{23}&-b_{23}& a_{22}&-b_{22}&-b_{21}&a_{21}\\
b_{23}&a_{23}& b_{22}& a_{22}& a_{21}&b_{21}\\
b_{13}&a_{13}& b_{12}& a_{12}& a_{11}&b_{11}\\
a_{13}&-b_{13}& a_{12}&-b_{12}&-b_{11}&a_{11}\\
\end{pmatrix},%\quad  \group{W}_{ij}=a_{ij}+ib_{ij}.
\end{equation}
where the entries are the real and imaginary parts of the elements
of the $SU(3)$-matrix, $\group{W}_{ij}=a_{ij}+ib_{ij}$.

Finally, to parametrize the elements of the coset space we calculate 
the remaining exponential map
\begin{equation}
\group{S}=\exp \left \lbrace
\algebra{s} \right\rbrace\quad\hbox{with}\quad
\algebra{s}=\sum_{i=1}^{i=6} s_i\algebra{u}_{8+i}\,.\label{cosetgen}
\end{equation}
The result depends on the real parameter $\sigma=\|\vec{s}\,\|$
and the $6$-dimensional unit-vector $\hat s=\vec{s}/\|\vec{s}\,\|$. In a
$1\times 6$-block notation the map takes the form
\begin{equation}
\group{S}=\begin{pmatrix}\cos 2\sigma&-\sin 2\sigma\,\hat s^\trnsp\cr \sin 2\sigma\,\hat s&\group{S}_\perp
 \end{pmatrix}\label{Smatrix}
\end{equation}
with $6$-dimensional matrix
\begin{equation}
\group{S}_\perp=\cos \sigma\,\id+\sin \sigma\,\hat{\algebra{s}}_\perp+\left(\cos 2\sigma-\cos \sigma\right)\hat s\hat s^\trnsp
+(1-\cos \sigma)\hat v\hat v^\trnsp\,.\label{chcoset}
\end{equation}
The matrix $\hat{\algebra{s}}_\perp$ is the $6\times 6$ right-lower block of $\algebra{s}$
in \eqref{cosetgen}.
The unit-vector $\hat v^\trnsp=(\hat s_2,-\hat s_1,\hat s_4,-\hat s_3,-\hat s_6,\hat s_5)$
defining the last projector in \eqref{chcoset} is orthogonal to the unit-vector
$\hat s$ defining the projector $\hat s\hat s^\trnsp$.

In the numerical integration we need the exponential map for elements 
$\algebra{u}$ in $\algebra{g}_2$. They are related to the generators used in the factorization
by the Baker-Campbell-Hausdoff formula,
\begin{equation}
\delta \tau \, \algebra{u}=\delta \tau
\left(\algebra{s}+\algebra{v}\right)+\frac{1}{2}\delta \tau^2
\komm{\algebra{s}}{\algebra{v}}+ \cdots\label{baker}
\end{equation}
Depending on the order of the symplectic integrator we must solve this relation
for $\algebra{s}$ and $\algebra{v}$ up to the corresponding order in $\delta \tau$. 
For a second order integrator used in this work this can be done analytically since the commutator
$[\algebra{s},\algebra{v}]$ does not contain any contribution
of the sub-algebra $\algebra{su}(3)$. The integrator used in the (L)HMC algorithm must be time 
reversible. It can be checked that time reversibility holds to every order in
this expansion. To summarize, for a second order integrator the approximation
\eqref{baker} may be used in the exponentiations needed to calculate
$\group{V}$ and $\group{S}$. This approximation leads to a violation of energy conservation 
which is of the same order as the violation one finds with a second order integrator. 
In comparison to the exponentiation via the spectral decomposition 
the method based on the factorization \eqref{groupdecomp} is more than ten
times faster. It is also much faster than computing the exponential map for $SO(7)$ via the 
Cayley-Hamilton theorem.

\subsection{Exponential error reduction for Wilson loops}
\noindent
In the confining phase the rectangular Wilson loop
scales as $W(L,T) \propto \exp(-\sigma L\cdot T)$. In order to estimate 
the string tension $\sigma$ we probe areas $LT$ ranging from $0$ up to 
$100$ and thus $W$ will vary by approximately $40$ orders of magnitude. 
A brute force approach where statistical errors for the expectation value
 of Wilson or Polyakov loops 
decrease with the inverse square root of the number of statistically 
independent configurations by just increasing the number of generated 
configurations will miserably fail. Nevertheless,
convincing results on $G_2$ Casimir scaling on intermediate scales
for representions with relative Casimirs $\cC'_\rep\leq 5$ have 
been obtained in \cite{Liptak:2008gx} with a variant of the 
smearing procedure. When reproducing these results we observed 
that the calculated string tensions depend sensitively on the smearing
parameter\footnote{This is not the case for the ratios of string tensions.}.
Thus to obtain accurate and reliable numbers for the static potential and to
detect string breaking we implemented the multi-step L\"uscher-Weisz 
algorithm with exponential error-reduction for the time transporters
of the Wilson-loops  \cite{Luscher:2001up}.
With this method the absolute errors of Wilson lines decrease
\\[.5mm]
\noindent
\begin{minipage}{7.4cm}
 exponentially 
with  the temporal extent  $T$ of the line. This is achieved by subdividing the 
lattice into $n_t$ sublattices  $V_1,\dots,V_{n_t}$
containing the Wilson loop and separated by time slices 
plus the remaining sublattice, denoted by $\bar V$,
see figure on the right. At the first level in a two-level algorithm 
the time extent of each sublattice $V_n$ is $4$ such that $n_t$ is the 
smallest natural number with $4n_t\geq T+2$.
In the figure on the right $T=14$ and the lattice is split into four
sublattices $V_1,V_2,V_3,V_4$ containing the Wilson loop plus
the complement $\bar V$. The Wilson loop is the product
of parallel transporters $W=\group{T}_2'\group{T}_3'\group{T}_4\group{T}_3 \group{T}_2 \group{T}_1$.
If a sublattice $V_n$ contains only one connected piece of the Wilson loop (as $V_1$ and $V_4$
do) then one needs to calculate the sublattice expectation value
\begin{equation}
\langle \group{T}_n\rangle_{n}=\frac{1}{Z_{n}}\int
\limits_\text{\text{sublattice} n} \mathcal{D}\gU\, \group{T}_n \,e^{-S},
\end{equation}
if $V_n$ contains two connected pieces (as $V_2$ and $V_3$)
\end{minipage}
\hfill
%\begin{figure}[htb]
\begin{minipage}{9.6cm}
%\begin{figure}[htb]
    \begin{tikzpicture}[scale=0.73]
\fill[gray,fill opacity=0.5] (0,1) rectangle (6,2);
\fill[gray,fill opacity=0.5] (0,10) rectangle (6,11);
\fill[red, fill opacity=0.5] (0,2) rectangle (6,4);
\fill[green, fill opacity=0.5] (0,4) rectangle (6,6);
\fill[blue, fill opacity=0.5] (0,6) rectangle (6,8);
\fill[red, fill opacity=0.5] (0,8) rectangle (6,10);
\draw (0,1) grid[step=0.5] (6,11);
\pgfsetlinewidth{2pt}
\draw[->] (1,4) -- (1,2.5) -- (5,2.5) -- (5,4);
\draw[->] (5,4) -- (5,6);
\draw[->] (5,6) -- (5,8);
\draw[->] (5,8) -- (5,9.5) -- (1,9.5) -- (1,8);
\draw[->] (1,8) -- (1,6);
\draw[->] (1,6) -- (1,4);
\path (3.5,1.85) node [below left,fill=white,fill
opacity=0.8,text opacity=1] {$\bar V$};
\path (3.5,3.5) node [below left,fill=white,fill
opacity=0.8,text opacity=1] {$V_1$};
\draw (1.35,2.75) node[right,fill=white,fill opacity=0.8,circle,inner sep=1pt]
{$\group{T}_1$};
\path (3.5,5.25) node [below left,fill=white,fill
opacity=0.8,text opacity=1] {$V_2$};
\draw (4.9,4.7) node[right,fill=white,fill opacity=0.8,circle,inner sep=1pt]
{$\group{T}_2$}; \draw (.1,4.7) node[right,fill=white,fill
opacity=0.8,circle,inner sep=1pt] {$\group{T}'_2$}; \path (3.5,7.25) node [below left,fill=white,fill
opacity=0.8,text opacity=0] {$V_3$};
\draw (4.9,6.7) node[right,fill=white,fill opacity=0.8,circle,inner sep=1pt]
{$\group{T}_3$}; \draw (.1,6.7) node[right,fill=white,fill
opacity=0.8,circle,inner sep=1pt] {$\group{T}'_3$}; \path (3.5,9.25) node [below left,fill=white,fill
opacity=0.8,text opacity=0] {$V_4$};
\draw (1.35,9.2) node[right,fill=white,fill opacity=0.8,circle,inner sep=1pt]
{$\group{T}_4$}; \path (3.5,10.95) node [below left,fill=white,fill
opacity=0.8,text opacity=1] {$\bar V$};
\end{tikzpicture}
%%%%%%%%%%%%%%%%%%%%%%%%%%
\hskip.5cm
    \begin{tikzpicture}[scale=0.73]
\fill[gray,fill opacity=0.5] (0,1) rectangle (6,2);
\fill[gray,fill opacity=0.5] (0,10) rectangle (6,11);
\fill[red, fill opacity=0.5] (0,2) rectangle (6,3);
\fill[red!50!green, fill opacity=0.5] (0,3) rectangle (6,4);
\fill[green, fill opacity=0.5] (0,4) rectangle (6,5);
\fill[green!50!blue, fill opacity=0.5] (0,5) rectangle (6,6);
\fill[blue, fill opacity=0.5] (0,6) rectangle (6,7);
\fill[blue!50!red, fill opacity=0.5] (0,7) rectangle (6,8);
\fill[red, fill opacity=0.5] (0,8) rectangle (6,9);
\fill[red!50!gray, fill opacity=0.5] (0,9) rectangle (6,10);
\draw (0,1) grid[step=0.5] (6,11);
\pgfsetlinewidth{2pt}
\draw[->] (1,3) -- (1,2.5) -- (5,2.5) -- (5,3);
\draw[->] (5,3) -- (5,4);
\draw[->] (5,4) -- (5,5);
\draw[->] (5,5) -- (5,6);
\draw[->] (5,6) -- (5,7);
\draw[->] (5,7) -- (5,8);
\draw[->] (5,8) -- (5,9);
\draw[->] (5,9) -- (5,9.5) -- (1,9.5) -- (1,9);
\draw[->] (1,9) -- (1,8);
\draw[->] (1,8) -- (1,7);
\draw[->] (1,7) -- (1,6);
\draw[->] (1,6) -- (1,5);
\draw[->] (1,5) -- (1,4);
\draw[->] (1,4) -- (1,3);
\path (3.4,1.9) node [below left,fill=white,fill
opacity=0.8,text opacity=1] {$\bar V$};
\path (3.5,2.9) node [below left,fill=white,fill
opacity=0.8,text opacity=1] {$V_{11}$};
\path (3.5,3.9) node [below left,fill=white,fill
opacity=0.8,text opacity=1] {$V_{12}$};
\path (3.5,4.9) node [below left,fill=white,fill
opacity=0.8,text opacity=1] {$V_{21}$};
\path (3.5,5.9) node [below left,fill=white,fill
opacity=0.8,text opacity=1] {$V_{22}$};
\path (3.5,6.9) node [below left,fill=white,fill
opacity=0.8,text opacity=0] {$V_{31}$};
\path (3.5,7.9) node [below left,fill=white,fill
opacity=0.8,text opacity=0] {$V_{32}$};
\path (3.5,8.9) node [below left,fill=white,fill
opacity=0.8,text opacity=0] {$V_{41}$};
\path (3.5,9.9) node [below left,fill=white,fill
opacity=0.8,text opacity=0] {$V_{42}$};
\path (3.4,10.9) node [below left,fill=white,fill
opacity=0.8,text opacity=1] {$\bar V$};
\end{tikzpicture}
%\end{figure}
\end{minipage}$\quad$\\[1mm]
then one needs to calculate
$\langle \group{T}_n\otimes \group{T}'_n\rangle_{n}$. The
updates in each sublattice are done with fixed link variables 
on the time-slices bounding  the sublattice. Calculating the expectation value of the 
full Wilson loop reduces to averaging over the links in the $n_t+1$ time slices,
\begin{equation}
 \langle W\rangle =\Big\langle\cC\Big( \langle \group{T}_1\rangle_1\langle \group{T}_2\otimes \group{T}_2'\rangle_{2}
\cdots \langle \group{T}_{n_t-1}\otimes \group{T}'_{n_t-1}\rangle_{n_t-1}\langle \group{T}_ {n_t}\rangle_{n_t}\Big)
\Big\rangle_\text{boundaries}\label{nested1}
\end{equation}
Here $\cC$ is that particular contraction of indices that leads to
the trace of the product $W=\group{T}'_2\cdots \group{T}'_{n_t-1}
\group{T}_{n_t}\group{T}_{n_t-1}\cdots \group{T}_2 \group{T}_1$.
In a two-level algorithm each sublattice $V_n$
is further divided into two sublattices $V_{n,1}$ and $V_{n,2}$, see right panel
in the above figure, and the sublattice 
updates are done on the small sublattices $V_{n,k}$ with fixed link variables on the
time slices separating 
the sublattices $V_{n,k}$. This way one finds two levels of nested averages. Iterating 
this procedure gives the \emph{multilevel algorithm}.
\\[1mm]
\begin{minipage}{12cm}
Since the dimensions $d_\rep$ grow rapidly
with the Dynkin labels $[p,q]$ -- for example, below we shall verify
Casimir scaling for charges in the $189$-dimensional representation $[2,1]$ --
it is difficult to store the many expectation values of tensor products
of parallel transporters. Thus we implemented a slight modification of the L\"uscher-Weisz
algorithm where the lattice is further split by a space slice with hyperplane
orthogonal to the plane defined by the Wilson loop, see figure on the right.
The sublattice updates are done with fixed link variables on the same time slices 
as before and in addition on the newly introduced space slice. Instead of $n_t$ sublattices 
containing the Wilson loop we now have $2n_t-2$ sublattices. But now every sublattice contains
only one connected part of the Wilson loop and \eqref{nested1} is replaced
\begin{equation}
 \langle W\rangle =\Big\langle \tr \prod_{n=1}^{2n_t-2}\langle \group{T}_n\rangle_n
\Big\rangle_\text{boundaries}\label{nested2}
\end{equation}
An iteration of this procedure by additional splittings of the time slices leads
again to a multilevel algorithm. 
In the present work we use a two level algorithm with time slices of length $4$
on the first and length $2$ on the second level. We calculate $\langle W\rangle$
for Wilson loops (and hence transporters $\group{T}_n$) of varying sizes
and in different representations. To avoid the storage
of tensor products of large representations we
implemented the modified algorithm as explained above. \\
\end{minipage}
\hfill
\begin{minipage}{5cm}
   \ \begin{tikzpicture}[scale=0.73]
\fill[gray,fill opacity=0.5] (0,1) rectangle (6,2);
\fill[gray,fill opacity=0.5] (0,10) rectangle (6,11);
\fill[red, fill opacity=0.5] (0,2) rectangle (6,3);
\fill[red!50!green, fill opacity=0.5] (0,3) rectangle (6,4);
\fill[green, fill opacity=0.5] (0,4) rectangle (6,5);
\fill[green!50!blue, fill opacity=0.5] (0,5) rectangle (6,6);
\fill[blue, fill opacity=0.5] (0,6) rectangle (6,7);
\fill[blue!50!red, fill opacity=0.5] (0,7) rectangle (6,8);
\fill[red, fill opacity=0.5] (0,8) rectangle (6,9);
\fill[red!50!gray, fill opacity=0.5] (0,9) rectangle (6,10);
\draw (0,1) grid[step=0.5] (6,11);
\pgfsetlinewidth{2pt}
\draw[->] (1,3) -- (1,2.5) -- (5,2.5) -- (5,3);
\draw[->] (5,3) -- (5,4);
\draw[->] (5,4) -- (5,5);
\draw[->] (5,5) -- (5,6);
\draw[->] (5,6) -- (5,7);
\draw[->] (5,7) -- (5,8);
\draw[->] (5,8) -- (5,9);
\draw[->] (5,9) -- (5,9.5) -- (1,9.5) -- (1,9);
\draw[->] (1,9) -- (1,8);
\draw[->] (1,8) -- (1,7);
\draw[->] (1,7) -- (1,6);
\draw[->] (1,6) -- (1,5);
\draw[->] (1,5) -- (1,4);
\draw[->] (1,4) -- (1,3);
\draw[-] (3,3) -- (3,9);
\path (3.4,1.9) node [below left,fill=white,fill
opacity=0.8,text opacity=1] {$\bar V$};
\path (3.5,2.9) node [below left,fill=white,fill
opacity=0.8,text opacity=1] {$V_{12}$};
\path (4.5,3.9) node [below left,fill=white,fill
opacity=0.8,text opacity=1] {$V_{13}$};
\path (4.5,4.9) node [below left,fill=white,fill
opacity=0.8,text opacity=1] {$V_{21}$};
\path (4.5,5.9) node [below left,fill=white,fill
opacity=0.8,text opacity=1] {$V_{22}$};
\path (4.5,6.9) node [below left,fill=white,fill
opacity=0.8,text opacity=0] {$V_{31}$};
\path (4.5,7.9) node [below left,fill=white,fill
opacity=0.8,text opacity=0] {$V_{32}$};
\path (4.5,8.9) node [below left,fill=white,fill
opacity=0.8,text opacity=0] {$V_{41}$};
\path (2.5,3.9) node [below left,fill=white,fill
opacity=0.8,text opacity=1] {$V_{11}$};
\path (2.5,4.9) node [below left,fill=white,fill
opacity=0.8,text opacity=1] {$V_{62}$};
\path (2.5,5.9) node [below left,fill=white,fill
opacity=0.8,text opacity=1] {$V_{61}$};
\path (2.5,6.9) node [below left,fill=white,fill
opacity=0.8,text opacity=0] {$V_{52}$};
\path (2.5,7.9) node [below left,fill=white,fill
opacity=0.8,text opacity=0] {$V_{51}$};
\path (2.5,8.9) node [below left,fill=white,fill
opacity=0.8,text opacity=0] {$V_{43}$};
\path (3.5,9.9) node [below left,fill=white,fill
opacity=0.8,text opacity=0] {$V_{42}$};
\path (3.4,10.9) node [below left,fill=white,fill
opacity=0.8,text opacity=1] {$\bar V$};
\end{tikzpicture}
\end{minipage}\\[-2mm]

We also applied the L\"uscher-Weisz algorithm to calculate the correlators of two
Polyakov loops $\langle P_\rep(0) P_\rep(R)\rangle$ on larger lattices. In this
case the complete lattice is divided into sublattices separated by time slices,
hence there is no complement $\bar V$. Since the Polyakov loops are only used
for lower-dimensional representations we have not split the lattice by a spatial
slicing but used tensor products similar to Eq.~\eqref{nested1}. Actually for the
calculations of Polyakov loop correlators we used the three-step L\"uscher-Weisz
algorithm.

%----------------------------------------------------------------------------
\section{String tension and Casimir scaling in $\bm{G_2}$ gluodynamics}
\noindent
The static inter-quark potential is linearly rising on intermediate distances
and the corresponding string tension will  depend on the representation of 
the static charges. We expect to find \emph{Casimir scaling} where the 
string tensions for different representations  $\rep$ and $\rep'$ scale according to
\begin{equation}
  \frac{\sigma_\rep}{c_\rep} = \frac{\sigma_{\rep'}}{c_{\rep'}}
\end{equation}
with quadratic Casimir $c_\rep$. Although all string tensions
will vanish at asymptotic scales it is still possible to check for Casimir scaling 
at intermediate scales where the linearity of the inter-quark potential is
nearly fulfilled.

To extract the static quark anti-quark potential two different methods are
available. The first makes use of the behavior of rectangular Wilson loops 
in representation $\rep$ for large $T$,
\begin{equation}
\vev{W_\rep(R,T)} = \exp\bigl( \kappa_\rep(R) - V_\rep(R)T
\bigr)\quad \text{with}\quad V_\rep(R) =
\gamma_\rep-\frac{\alpha_\rep}{R}+\sigma_\rep R.\label{pot}
\end{equation}
The potential can be extracted from the ratio of two Wilson loops
with different time-extent according to
\begin{equation}
V_\rep(R) = \frac{1}{\tau} \ln\frac{\vev{W_\rep(R,T)}}{\vev{W_\rep(R,T+\tau)}}.
\label{pot3}
\end{equation}
We calculated the expectation values of Wilson loops with the two-level
L\"uscher-Weisz algorithm and fitted the right hand side of \eqref{pot3} with
the potential $V_\rep(R)$ in \eqref{pot}. The fitting has been done for
external charges separated by one lattice unit up to separations $R$ with acceptable
signal to noise ratios. From the fits we extracted the constants
$\gamma_\rep,\alpha_\rep$ and $\sigma_\rep$ entering the static potential.
For an easier comparison of the numerical results on lattices of
different size and for different values of $\beta$ we subtracted the constant 
contribution to the potentials and plotted
\begin{equation}
 \tilde V_\rep(R)=V_\rep(R)-\gamma_\rep\label{potsubtract}
\end{equation}
in the figures. The statistical errors are determined with the Jackknife method.
In addition we determined the \emph{local string tension} 
\begin{equation}
\sigma_{\rm loc,\rep}\left(R+\frac{\rho}{2}\right)
=\frac{V_\rep(R+\rho)-V_\rep(R)}{\rho},\label{locsigma1}
\end{equation} 
given by the Creutz ratio
\begin{equation}
\sigma_{\rm loc,\rep}\left(R+\frac{\rho}{2}\right)=
\frac{1}{\tau\rho}\ln\frac{\vev{W_\rep(R+\rho,T)}
\vev{W_\rep(R,T+\tau)}}{\vev{W_\rep(R+\rho,T+\tau)} \vev{W_\rep(R,T)}}=
\frac{\alpha_\rep}{R(R+\rho)}+\sigma_\rep .\label{locsigma2}
\end{equation}

The second method to calculate the string tensions uses correlators of two Polyakov loops,
\begin{equation}
V_\rep(R) = -\frac{1}{\beta_T} \ln \vev{P_\rep(0) P_\rep(R)}.
\end{equation}
The correlators are calculated with the three-level L\"uscher-Weisz
algorithm and are fitted with the static potential $V_\rep(R)$ with fit
parameters $\gamma_\rep,\alpha_\rep$ and $\sigma_\rep$. 
Now the local string tension takes the form
\begin{equation}
\sigma_{\text{loc},\rep}\left(R+\frac{\rho}{2}\right) =
-\frac{1}{\beta_T\rho}\ln\frac{\vev{P_\rep(0) P_\rep(R+\rho)}}{\vev{P_\rep(0)
P_\rep(R)}}.
\end{equation}

\subsection{Casimir scaling in 3 dimensions}
\begin{figure}[htb]
\input{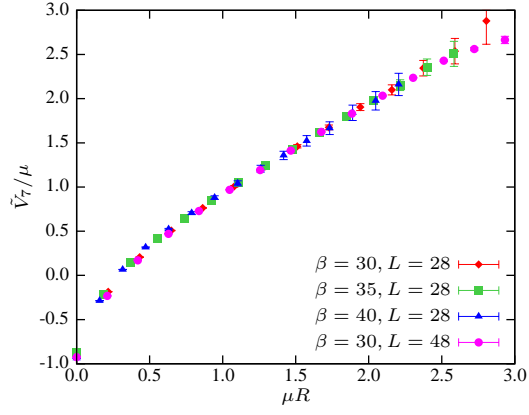}
\caption{Continuum scaling of the fundamental potential.}
\label{fig:contscal3d}
\end{figure}

\noindent
Most LHMC simulations are performed on a $28^3$ lattice with Wilson loops of
time-extent $T=12$.  To extract the static potentials from the ratio
of Wilson loops in \eqref{pot3} we chose $\tau=2$.
The fits to the static potential \eqref{pot} for charges in the
\emph{fundamental $7$-representation} and for values $\beta=30, 35$ and  $40$ yield the
lattice parameters $\alpha,\gamma$ and $\sigma$ given in Tab.~\ref{table3a}.
To check for scaling we plotted the potentials in `physical' units, $V/\mu$, with
mass scale set by the string tension in the $7$-representation,
\begin{equation}
\mu=\sqrt{\sigma_7},\label{units}
\end{equation}
as function of $\mu R$ in Fig. \ref{fig:contscal3d}.
We observe that the potentials for the three values of $\beta$ are the same
within error bars. In addition they agree with the potential (in physical units) 
extracted from the Polyakov loop on a much larger $48^3$ lattice.
\begin{table}
\caption{\label{tab:Potential} Potential for charges in the $7$-representation.}
\begin{ruledtabular}
\begin{tabular}{lcccc}
 & $\beta=30$, $L=28$ & $\beta=35$, $L=28$ & $\beta=40$, $L=28$ & $\beta=30$,
$L=48$  \\ \hline $\gamma a$ & $0.185(8)$ & $0.160(4)$ & $0.147(5)$ & $0.197(1)$ \\ 
 $\alpha$ & $0.0881(7)$ & $0.0752(3)$ & $0.071(4)$ & $0.098(1)$  \\
$\sigma a^2$ & $0.046(1)$ & $0.0340(8)$ & $0.024(1)$ & $0.0435(3)$ \\
\end{tabular}
\label{table3a}
\end{ruledtabular}
\end{table}

The fitted constants $\alpha_\rep,\gamma_\rep$ and $\sigma_\rep$
of the potential \eqref{pot} for the eight smallest representations are given in 
Tab.~\ref{tab:Potential1}. The Casimir scaling of coefficients becomes apparent
when they are divided by the corresponding coefficients of the static potential
in the $7$-representation. 
\begin{table}
\caption{\label{tab:Potential1} Fit-parameters of static potentials.}
\begin{ruledtabular}
\begin{tabular}{lcccccccc}
$\rep$ & $7$ & $14$ & $27$ & $64$ & $77$ &
$77'$ & $182$ & $189$ \\ \hline 
$\gamma_\rep a$ & $0.147(5)$ & $0.29(1)$ & $0.34(1)$ & $0.51(1)$ & $0.58(1)$ &
$0.74(2)$ & $0.83(1)$ & $0.77(2)$\\ 
$\gamma_\rep a/\cC'_\rep$ & $0.147$ & $  0.145$ & $0.146$ & $0.146$ &
$0.145$ & $0.148$ & $0.138$ &$0.144$\\ 
$\gamma_\rep/\gamma_{7}$ & $1$ & $1.97$ & $2.31$ & $3.46$ & $3.94$ &
$5.03$ & $5.64$ & $5.23$\\ 
$\alpha_\rep$ & $0.071(4)$ & $0.145(8)$ & $0.16(1)$ & $0.24(1)$  & $0.27(1)$ &
$0.36(1)$ &$0.37(1)$ & $0.36(1)$ \\
$\alpha_\rep/\cC'_\rep $ & $0.071$ & $0.0725$ & $0.069$ & $0.069$  & $0.068$ &
$0.072$ &$0.062$ & $0.068$ \\
$\alpha_\rep/\alpha_{7}$ & $1$ & $2.04$ & $2.25$ & $3.38$  & $3.80$ &
$5.07$ &$5.21$ & $5.07$ \\
$\sigma_\rep a^2$ & $0.024(1)$ & $0.048(2)$ & $0.057(3)$ &
$0.086(4)$ & $0.099(5)$ & $0.120(6)$ & $0.157(6)$ & $0.132(6)$\\
$\sigma_\rep a^2/\cC'_\rep$ & $0.024$ & $0.024$ & $0.024$ &
$0.025$ & $0.025$ & $0.024$ & $0.026$ & $0.025$\\
$\sigma_\rep/\sigma_{7}$ & $1$ & $2.00$ & $2.37$ &
$3.58$ & $4.12$ & $5.00$ & $6.54$ & $5.50$
\end{tabular}
\end{ruledtabular}
\end{table}

The \emph{local string tensions} extracted from the Creutz ratio can be determined 
much more accurately as the global string tensions extracted from
fits to the static potentials. Tab.~\ref{tab:Potential3} contains the local
string tensions for static charges in the eight smallest representations for $\rho=1$ and
different $R$ in \eqref{locsigma2}, divided by the corresponding
local string tensions in the $7$-representation. The results are insensitive
the the distance $R$ in the Creutz ratio. They agree within $1$ percent with the
values for the Casimir ratios $\cC'_\rep=\cC_\rep/\cC_7$ given in the last row of that table.
\begin{table}
\caption{\label{tab:Potential3} Scaled local string tension.}
\begin{ruledtabular}
\begin{tabular}{lcccccccc}
$\rep$ & $7$ & $14$ & $27$ & $64$ & $77$ &
$77'$ & $182$ & $189$ \\ \hline 
$\sigma_\rep(1/2)/\sigma_{7}(1/2)$ & $1$ & $1.9996(3)$ & $2.3327(5)$ &
$3.4981$ & $3.997(2)$ & $4.996(3)$ & $5.991(5)$ & $5.328(4)$\\ 
$\sigma_\rep(3/2)/\sigma_{7}(3/2)$ & $1$ & $1.99897$ & $2.3311$ &
$3.495(5)$  & $3.994(4)$ & $4.9897$ &$5.991$ & $5.321(9)$ \\ 
$\sigma_\rep(5/2)/\sigma_{7}(5/2)$ & $1$ & $1.9961$ & $2.3271$ &
$3.484(5)$ & $3.9807$ & $4.961$ & $5.94(2)$ & $5.291$\\
$\cC'_\rep$ & $1$ & $2.0000$ & $2.3333$ & $3.5000$ & $4.0000$ & $5.0000$ & $6.0000$ &
$5.333$\\
\end{tabular}
\end{ruledtabular}
\end{table}

\begin{figure}
\input{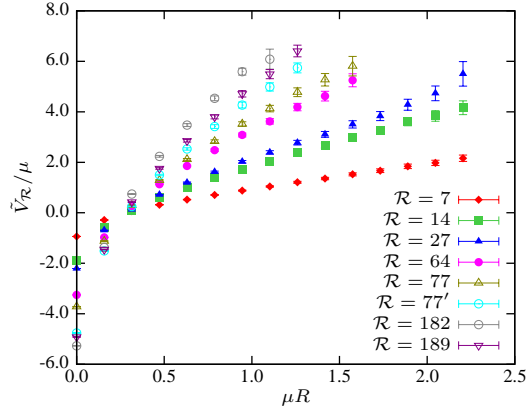}
\caption{Unscaled potential with $\beta=40$ on a $28^3$ lattice.}
\label{fig:potential40unscaled3d}
\end{figure}

In Fig.~\ref{fig:potential40unscaled3d} we plotted the
values for the eight potentials $V_7,\dots,V_{189}$ (with statistical errors)
measured in `physical units' $\mu$ defined in \eqref{units}.
The distance of the charges is measured in the same system of units. 
The linear rise at intermediate scales is 
clearly visible, even for charges in the $189$-dimensional 
representation.

\begin{figure}
\input{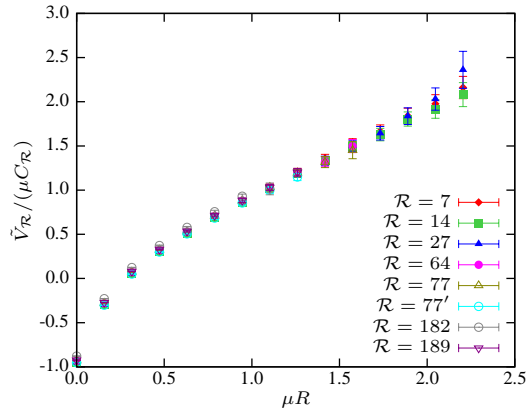}
\caption{Scaled potential with $\beta=40$ on a $28^3$ lattice.}
\label{fig:potential40scaled3d}
\end{figure}

Fig.~\ref{fig:potential40scaled3d} contains the
same data points rescaled with the quadratic Casimirs of the
corresponding representations. The eight rescaled potentials fall
on top of each other within error bars. This implies that
the \emph{full potentials} for short and intermediate separations 
of the static charges show Casimir scaling.

To further check for Casimir scaling we calculated the \emph{local string tensions}
with $\rho=1$, this time for all $R$ between $1$ and $10$ and 
not only for $R=0,1,2$ as in Tab.~\ref{tab:Potential3}. 
The horizontal lines are the values predicted by the Casimir scaling 
hypothesis. Clearly we see no sign of Casimir scaling violation on a $28^3$-lattice
near the continuum at $\beta=40$. Of course, for widely separated charges 
in higher  dimensional representations the error bars are not negligible even 
for an algorithm with exponential error reduction.

\begin{figure}
\input{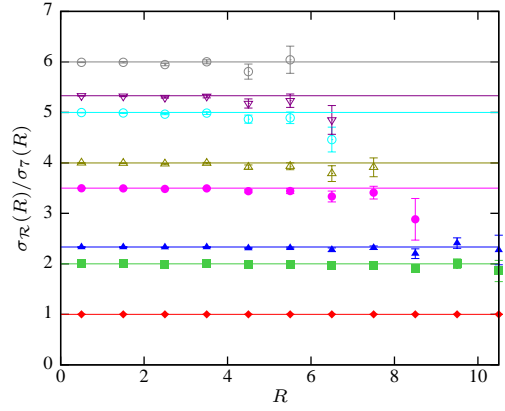}
\caption{Ratio of the local string Tension with $\beta=40$ scaled on a $28^3$
lattice for the eight smallest representations.}
\label{fig:stringtension40scaled3d}
\end{figure}

\subsection{L\"uscher term}
\begin{figure}[htb]
\input{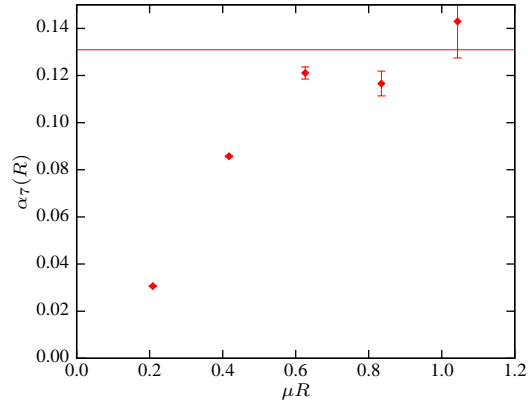}
\caption{Local L\"uscher term on a $48^3$ lattice at $\beta=30$.}
\label{fig:luescher}
\end{figure}

\noindent
In Tab.~\ref{tab:Potential1} we have seen that the dimensionless coefficient 
$\alpha_\rep$ in the static potential  scales with the quadratic Casimir, 
similarly to the string tension. The corresponding term, if measured at distances where 
the flux tube has already formed, is referred to as \emph{L\"uscher term}. Its value has been 
calculated by L\"uscher for charges in the fundamental representation, 
in $d$ dimensions $\alpha=(d-2)\pi/24$, and it is believed to 
be universal \cite{Luscher:1981}. The value  $\alpha=\pi/24$ in $3$ dimensions 
is off the results in Tab.~\ref{tab:Potential}. However, since the
coefficients in this table are fitted to the static potential from $R=1$ to values of $R$ with
acceptable signal to noise ratio, they contain contributions from the
short range Coulombic tail. To calculate $\alpha_\rep$ at intermediate distances
we better use the (local) L\"uscher term
\begin{equation}
\alpha_{\rm loc,\rep}\left(R\right)=
\frac{R^3}{2\beta_T\rho^2}\ln\frac{\vev{P_\rep(0) P_\rep(R+\rho)}\vev{P_\rep(0)
P_\rep(R-\rho)}}{\vev{P_\rep(0) P_\rep(R)}\vev{P_\rep(0) P_\rep(R)}}=
\frac{\alpha_\rep R^2}{R^2-\rho^2} ,\label{ratioluescher2}
\end{equation}
with  $\rho=1$. In Fig.~\ref{fig:luescher} we plotted the local L\"uscher term  
for charges in the $7$-representation on a larger $48^3$-lattice with  $\beta=30$. 
Our data at intermediate distances  are in agreement with the theoretical 
prediction $\alpha_7=\pi/24\approx 0.131.$

\subsection{String breaking and glue-lumps in 3 dimensions}
\begin{figure}
\input{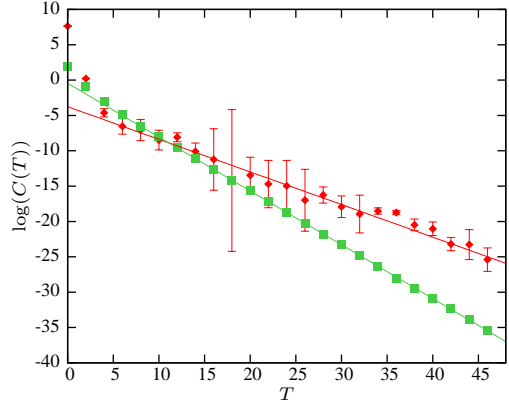}
\caption{Glue-lump correlator (lattice size $48^3$, $\beta = 30$).}
\label{fig:gluelumpmass1}
\end{figure}

\begin{figure}[htb]
\input{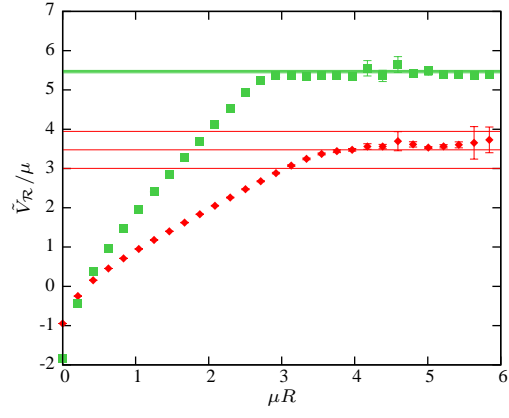}
\caption{Potential for both fundamental representations (lattice size $48^3$,
$\beta = 30$) and corresponding glue-lump mass.}
\label{fig:gluelumpmass2}
\end{figure}

\noindent
To observe the breaking of strings connecting static charges
at intermediate scales when one further increases the separation
of the charges we performed high statistics LHMC simulations on a $48^3$ 
lattice with $\beta=30$. We calculated  expectation values of Wilson loops 
and products of Polyakov loops for charges in the two fundamental representations 
of $G_2$. When a string breaks then each  static charge in the representation
$\rep$ at the end of the string is screened by $N(\rep)$ gluons to form a 
colour blind glue lump. We expect that the dominant decay channel for an 
over-stretched string is string $\to$ gluelump~$+$~gluelump. For a string to
decay the energy stored in the string must be sufficient to produce two glue-lumps.
According to \eqref{eq:representationsG2p} it requires at least $3$ gluons
to screen a static charge in the $7$-representation, one gluon to screen
a charge  in the $14$-representation and two gluons to screen a charge 
in the  $27$-representation. 
We shall calculate the separations of the charges where string breaking
sets in and the masses of the produced glue-lumps. The mass of such
a quark-gluon bound state can be obtained from the correlation function
\begin{equation}
C_\rep (T)=\erw{\left. \left(\bigotimes \limits_{n=1}^{N(\rep)}F_{\mu\nu}(y)\right)\right \vert_{\rep,a} \rep(\gU_{yx})_{ab} \left. \left(\bigotimes \limits_{n=1}^{N(\rep)}F_{\mu\nu}(x)\right)\right \vert_{\rep,b}} \propto \exp{\left(-m_\rep T\right)},\label{corrlumps}
\end{equation}
where $\rep(\gU_{yx})$ is the temporal parallel transporter in the representation
$\rep$ from $x$ to  $y$ of length $T$.  It represents 
the static sources in the representation  $\rep$ . The vertical line means projection 
of the tensor product onto that linear subspace on which the irreducible
representation $\rep$ acts,
\begin{equation}
\left(14\otimes 14\otimes\cdots\otimes 14\right)=\rep\oplus \cdots\;.
\end{equation}
For example, for charges in the $14$-representation the projection is simply
\begin{eqnarray}
F_{\mu\nu}(x)\Big\vert_{14,a}=F^a_{\mu\nu}(x),\quad\hbox{where}\quad 
F_{\mu\nu}^a T^a =F_{\mu\nu}.
\end{eqnarray}
For charges in the $7$-representation we must project the reducible
representation $14\otimes 14\otimes 14$ onto the irreducible representation $7$. 
Using the embedding of $G_2$ into $SO(7)$ representations  one shows that this 
projection can be done with the help of the totally antisymmetric $\varepsilon$-tensor 
with $7$ indices,
\begin{eqnarray}
F_{\mu\nu}(x)\otimes F_{\mu\nu}(x) \otimes F_{\mu\nu}(x)\Big\vert_{7,a}\propto
 F^p_{\mu\nu} (x)F^q_{\mu\nu} (x) F^r_{\mu\nu} (x)\varepsilon_{a bc de fg} 
T^p_{bc}T^q_{de} T^r_{fg}.
\end{eqnarray}
Fig.~\ref{fig:gluelumpmass1} shows the logarithm of the glue-lump correlator
\eqref{corrlumps} as function of the separation of the two lumps for static
charges in the fundamental representations $7$ and $14$.
 The linear fits to the data yield the glue-lump masses
\begin{equation}
m_7=0.46(4), \quad m_{14}=0.767(5).
\end{equation}
Thus we expect that the subtracted static potentials approach the
asymptotic values
\begin{equation}
\tilde V_\rep\longrightarrow
2m_\rep-\gamma_\rep.
\end{equation}
With the fit-values $\gamma_7=0.197(1)$ and $\gamma_{14}=0.381(2)$ we find 
\begin{equation}
\tilde V_7/\mu\longrightarrow 3.46\quad,\quad
\tilde V_{14}/\mu \longrightarrow 5.52.\label{asvalues}
\end{equation}
Fig.~\ref{fig:gluelumpmass2} shows the rescaled potentials for charges
in the fundamental representations together with the asymptotic
values \eqref{asvalues} extracted from the glue-lump correlators.
Within error bars both potentials flatten exactly at separations of
the charges where the energy stored in the flux tube is twice 
the glue-lump energy.

\begin{figure}[htb]
\input{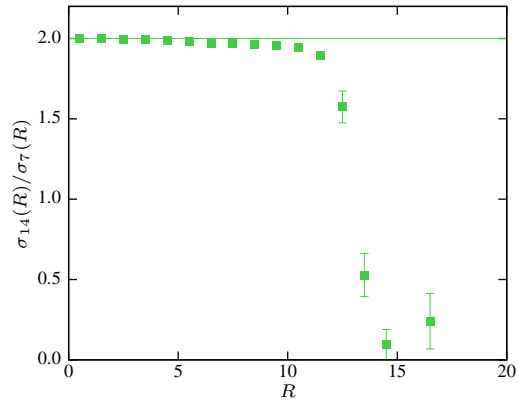}
\caption{Casimir scaling of local string tension ($48^3$ lattice, $\beta
= 30$).}
\label{fig:gluelumpmass4}
\end{figure}

A good approximation for the string breaking distance  is then given by $V_\rep(R^\text{cr}) \approx 2\,m_{\rep}$.
Assuming Casimir scaling for the coefficients $\alpha_\rep,\gamma_\rep$ and $\sigma_\rep$
in the static potential we obtain
\begin{equation}
\mu R_{\rep}^{\text{cr}} =
\left(\sqrt{ \alpha_{7}+\frac{1}{4}\left(\frac{\gamma_{7}}{\mu}
-M_\rep\right)^2}-\frac{1}{2}
\left(\frac{\gamma_{7}}{\mu}-M_\rep\right)\right),\quad M_\rep=\frac{2m_{\rep}}{\mu\cC'_\rep}.
\label{Rcrit}
\end{equation}
Inserting the result from the last row in Tab.~\ref{tab:Potential} and
the glue-lump masses we find $\mu R^{\text{cr}}_7=4.00$ and
$\mu R^{\text{cr}}_{14}=3.28$. These values agree well with
the separations $\mu R$ in Fig.~\ref{fig:gluelumpmass2} where the
static potentials flatten such that string breaking
sets in at scales predicted by formula \eqref{Rcrit}.
\begin{figure}[htb]
\input{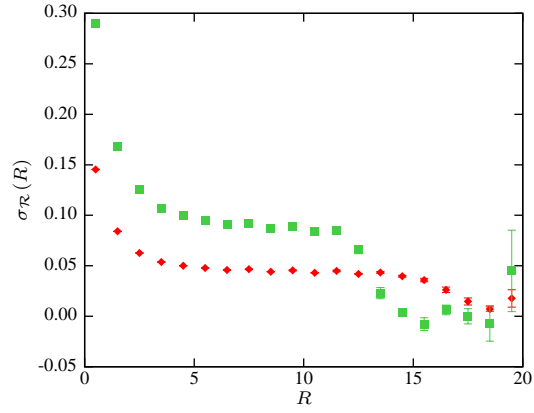}
\caption{Local string tension ($48^3$ lattice, $\beta = 30$).}
\label{fig:gluelumpmass3}
\end{figure}
Fig.~\ref{fig:gluelumpmass3} shows the local string tensions in the 
two fundamental representation and  Fig.~\ref{fig:gluelumpmass4} their
ratios. Especially the last plot makes clear that the string
connecting charges in the adjoint representation break earlier 
than the string connecting charges in the $7$-representation.
The formula \eqref{Rcrit} predicts $R^{\text{cr}}_{14}=9.40$
and just above this separation the ratio of local string tensions 
$\sigma_{14}(R)/\sigma_7(R)$ shows indeed a pronounced knee.

\FloatBarrier

\subsection{Casimir scaling in 4 dimensions}
\noindent
In this last section we present our results for the static potential 
in $4$ dimensions. The local HMC-simulations have been performed on a 
small $14^4$ and a larger $20^4$ lattice for different values of $\beta$.
The static potentials and local string tensions have been extracted
from \eqref{pot3} and \eqref{locsigma2}, where  the expectation values
have been calculated with a two-step  L\"uscher-Weisz algorithm.
Tab.~\ref{tab:Potential4D} contains the fits to the parameters in
the potential for static charges in the $7$-representation for these 
lattices and values for $\beta$.
\begin{table}
\caption{\label{tab:Potential4D} Parameters of the quark anti-quark 
potential in $4$ dimensions.}
\begin{ruledtabular}
\begin{tabular}{lccc}
 & $\beta=9.7$, $L=14$ & $\beta=10$, $L=14$ & $\beta=9.7$, $L=20$ 
\\ \hline $\gamma a$ & $0.83(8)$ & $0.74(4)$ & $0.68(9)$  \\ 
 $\alpha$ & $0.40(7)$ & $0.33(3)$ & $0.28(8)$  \\
$\sigma a^2$ & $0.07(2)$ & $0.042(9)$  & $0.11(1)$ \\
\end{tabular}
\label{table3b}
\end{ruledtabular}
\end{table}

\begin{figure}[htb]
\input{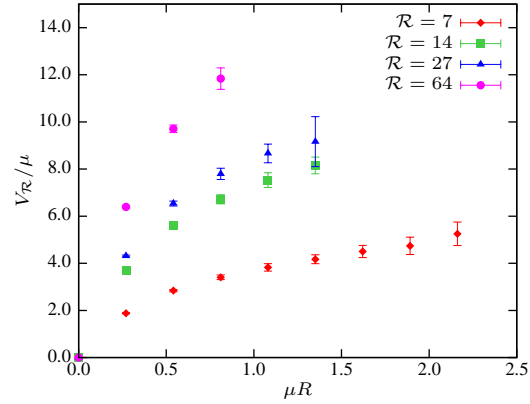}
\caption{Unscaled potential at $\beta=9.7$ on a $14^4$ lattice.}
\label{fig:potential97unscaled4d}
\end{figure}

\begin{figure}[htb]
\input{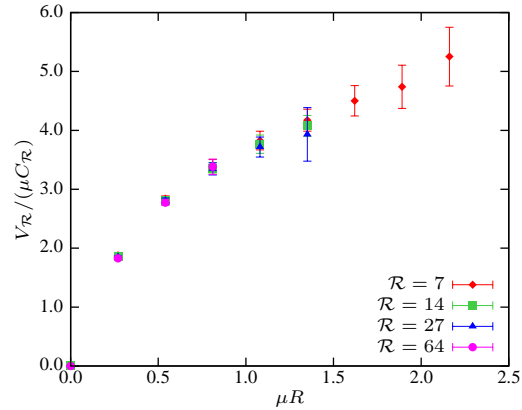}
\caption{Scaled potential at $\beta=9.7$ on a $14^4$ lattice.}
\label{fig:potential97scaled4d}
\end{figure}

Fig.~\ref{fig:potential97unscaled4d} shows the static potentials in
`physical units' $\mu=\sqrt{\sigma_7}$ for charges in the $7,14,27$ and
$64$-dimensional representations and coupling $\beta=9.7$ as function of the distance
between the charges in physical units. The corresponding value for $\sigma_7$ 
is taken from Tab.~\ref{tab:Potential4D}. The same coupling has been 
used in \cite{Liptak:2008gx} on an asymmetric $14^3\times 28$ lattice.
After normalizing the potential with the quadratic Casimirs they
are identical within error bars, as can be seen in
Fig.~\ref{fig:potential97scaled4d}. Our findings are in complete agreement with
the results in \cite{Liptak:2008gx} on  Casimir scaling  in $4$-dimensional $G_2$-gluodynamics at $\beta=9.7$ 
and our accurate results on Casimir scaling on intermediate scales in $3$-dimensional 
$G_2$-gluodynamics.

\begin{figure}[htb]
\input{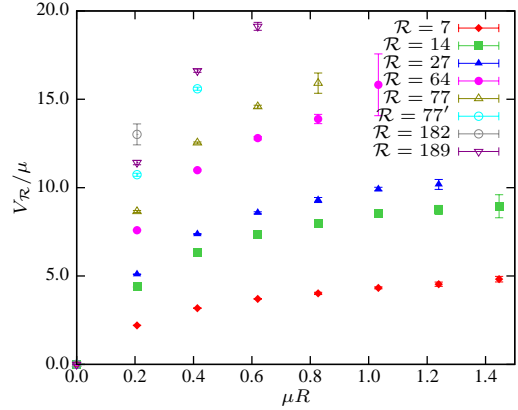}
\caption{Unscaled potential at $\beta=10$ on a $14^4$ lattice.}
\label{fig:potential10unscaled4d}
\end{figure}

\begin{figure}[htb]
\input{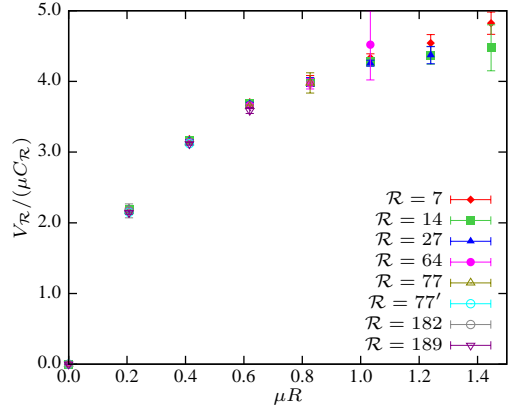}
\caption{Scaled potential at $\beta=10$ on a $14^4$ lattice.}
\label{fig:potential10scaled4d}
\end{figure}

Figs.~\ref{fig:potential10unscaled4d} and \ref{fig:potential10scaled4d} show the
corresponding results for a weaker coupling  $\beta=10$ closer to the continuum limit. For this small
coupling we can measure the potential only up to separations $\mu R\approx 1.5$
of the charges. But we can do this with high precision and for
higher-dimensional representations. As for $\beta=9.7$ we find that the 
potentials normalized with the second  order Casimirs fall on top of each other.
This confirms Casimir scaling for $G_2$-gluodynamics in $4$ dimensions
for charges in representations with dimensions $7,14,27,64,77,77',182$ and $189$.

\begin{figure}[htb]
\input{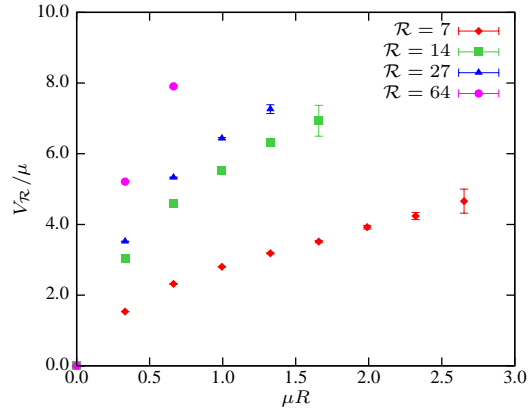}
\caption{Unscaled potential at $\beta=9.7$ on a $20^4$ lattice.}
\label{fig:potential97unscaled4dp}
\end{figure}

\begin{figure}[htb]
\input{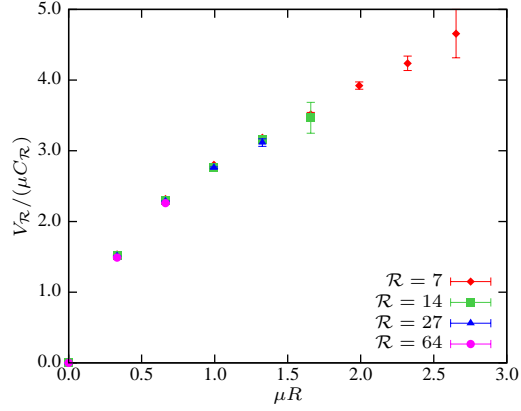}
\caption{Scaled potential at $\beta=9.7$ on a $20^4$ lattice.}
\label{fig:potential97scaled4dp}
\end{figure}

Finally we simulated on a much larger $20^4$ lattice at $\beta=9.7$ in order to
calculate the static potential for larger separations of the static quarks.
Unfortunately the distance $\mu R\approx 3$ is still not sufficient to detect
string breaking, see Fig.~\ref{fig:potential97unscaled4dp}. But again the
potentials normalized with the quadratic Casimirs shown in
Fig.~\ref{fig:potential97scaled4dp} are equal within error bars.

In Tab.~\ref{tab:Potential4D201} we have listed the fit-values for the parameters
of the potentials on the larger $20^4$ lattice for static charges in the
representations with dimensions $7,14$ and $27$. For all representation we find
Casimir scaling of all three parameters in the potential. Unfortunately the
fit-parameters cannot be determined reliably in the $64$-representation with the
present data. This is attributed to larger errors for the potentials at
intermediate scales, see Fig.~\ref{fig:potential97unscaled4dp}, so that the
parameters can only be determined from the ultraviolet part of the potential for
this representation ($R<3$) which is rather Coulomb-like than linearly rising.
Much more conclusive are the local string tensions calculated on the larger
lattice (now up to the $64$-representation). Tab.~\ref{tab:Potential4d3} contains
the local string tensions divided by the local string tensions in the
$7$-representation. These normalised values are constant up to separations of the
charges where the statistical errors are under control. Compared to the
corresponding numbers in $3$ dimensions, see Tab.~\ref{tab:Potential3}, we now see
a slight dependence of the local string tensions from Eq.~\ref{locsigma1} on the
distance $R$. Despite of the lower precision of the results in $4$ dimensions compared to the corresponding
results in $3$ dimensions we again confirm Casimir scaling on intermediate scales
within $5$ percent.

\begin{table}
\caption{\label{tab:Potential4D201} Fit-parameters of static potentials ($20^4$
lattice, $\beta=9.7$).}
\begin{ruledtabular}
\begin{tabular}{lccc}
$\rep$ & $7$ & $14$ & $27$ \\ \hline 
$\gamma_\rep a$ & $0.68(9)$ & $1.39(4)$ & $1.61(3)$  \\ 
$\gamma_\rep a/\cC'_\rep$ & $0.68$ & $  0.695$ & $0.690$ \\ 
$\alpha_\rep$ & $0.28(8)$ & $0.60(2)$ & $0.69(2)$   \\
$\alpha_\rep/\cC'_\rep $ & $0.28$ & $0.30$ & $0.295$  \\ 
$\sigma_\rep a^2$ & $0.11(1)$ & $0.21(1)$ & $0.251(9)$  \\
$\sigma_\rep a^2/\cC'_\rep$ & $0.11$ & $0.105$ & $0.107$  \\
\end{tabular}
\end{ruledtabular}
\end{table}

\begin{table}
\caption{\label{tab:Potential4d3} Scaled local string tension ($20^4$ lattice,
$\beta=9.7$).}
\begin{ruledtabular}
\begin{tabular}{lcccc}
$\rep$ & $7$ & $14$ & $27$ & $64$ \\ \hline 
$\sigma_\rep(1/2)/\sigma_{7}(1/2)$ & $1$ & $1.973(1)$ & $2.294(1)$ &
$3.396(8)$ \\ 
$\sigma_\rep(3/2)/\sigma_{7}(3/2)$ & $1$ & $1.987(3)$ & $2.303(4)$ &
$3.44(2)$  \\ 
$\sigma_\rep(5/2)/\sigma_{7}(5/2)$ & $1$ & $1.92(1)$ & $2.28(3)$ &
--- \\
$\cC'_\rep$ & $1$ & $2.0000$ & $2.3333$ & $3.5000$ \\
\end{tabular}
\end{ruledtabular}
\end{table}

\begin{figure}[htb]
\input{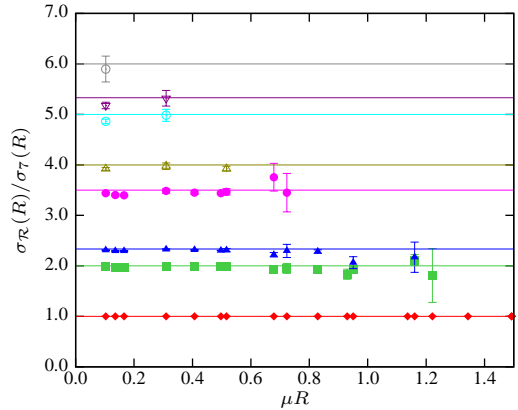}
\caption{Scaled local string Tension with $\beta=9.7,10$ on $14^4,20^4$
lattices.}
\label{fig:stringtensionscaled4d}
\end{figure}

All our simulation results for the local string tensions $\sigma_\rep(R)$ 
normalized by $\sigma_7(R)$ on a $14^4$-lattice with $\beta\in\{9.7,\,10\}$ and
on a $20^4$-lattice with $\beta=9.7$ and for $\mu R\leq 1.5$ 
are collected in  Fig.~\ref{fig:stringtensionscaled4d}. 
The horizontal lines in this figure
show the prediction of the Casimir scaling hypothesis. The normalized
data points are compatible with each other and with the hypothesis.

%============================================================================

\section{Conclusions}
\label{sect:conclusions}
\noindent
In the present work we implemented an efficient and fast LHMC algorithm 
to simulate $G_2$ gauge theory in three and four dimensions. With only a
slight modification we can include a (normalized) Higgs field in
the $7$-representation.  The corresponding results for the phase diagram
of $G_2$-Yang-Mills-Higgs theory will soon be presented in a companion
paper. The algorithm has been optimized with the help of the coset 
decomposition of group elements and the analytic expressions 
for the  exponential maps for the two factors. In addition we implemented
a slightly modified L\"uscher-Weisz multi-step algorithm with exponential error
reduction to measure the static potentials for charges in various $G_2$-representations. 
The accurate results in $3$ dimensions show that all parameters of
the fitted static potentials show Casimir scaling, see Tab.~\ref{tab:Potential}.
The global string tensions extracted from these fits 
show that possible deviations from Casimir scaling, if they exist, 
must be less than $4$ percent. We also extracted the local string tensions 
from the Creutz ratios to obtain even more precise data. This way
we confirm Casimir scaling with $1$ percent accuracy. 
Thus we conclude that in $3$-dimensional $G_2$-Gluodynamics the
string tensions show Casimir scaling for all charges in the representations 
with dimensions $7,14,27,64,77,77',182$ and $189$. 
In passing we can check the scaling formula \eqref{Nair} for the string tension
$\sigma_\rep(\beta)$ as function of the coupling $\beta\propto 1/g^2$ \cite{Karabali:2009rg}.
On a fixed lattice this formula implies that the product $\beta^2\sigma_\rep(\beta)$
should be independent of $\beta$. Using the values for the string tension $\sigma_7$
in Tab.~\ref{tab:Potential} we obtain
%\begin{tabular}
\begin{center}
\begin{minipage}{0.4\textwidth}
\begin{ruledtabular}
\begin{tabular}{lccc}
 & $\beta=30$ & $\beta=35$ & $\beta=40$ \\ \hline
$\sigma_7$ & $0.046(1)$ & $0.0340(8)$ & $0.024(1)$ \\
$\beta^2\sigma_7$ & $41.4$ & $41.7$ & $38.4$\\
\end{tabular}
\end{ruledtabular}
\end{minipage}
\end{center}
The numbers in the last row show that the scaling $\sigma\propto 1/\beta^2$
is almost fulfilled. In the present work we did not attempt to further clarify this 
interesting point by simulating at many $\beta$-values and using the more accurate 
local string tensions.

For charges in the two fundamental representations we performed LHMC simulations on
larger lattices to detect string breaking at asymptotic scales. In $3$ dimensions we 
observe that string breaking indeed sets in at the expected scale where the energy stored
in the flux tube is sufficient to create two glue lumps. To confirm this
expectation we calculated masses of glue lumps associated with
static charges in the fundamental representations. 
In $4$-dimensional $G_2$-gluodynamics we found Casimir scaling for charges in the representations
$7,14,27$ and $64$, similarly as we did 
in $3$ dimensions, although the uncertainties are of course larger. 
But within error bars we see no violation of Casimir scaling and
this confirms the corresponding results in \cite{Liptak:2008gx}, obtained with a variant 
of the smearing procedure. To see the expected string breaking in $4$ dimensions
one would need larger lattices than those used in the present work.
%============================================
\vspace{-0.4em}
\begin{acknowledgments}
\vspace{-0.4em}
\noindent
Helpful discussions with Philippe de Forcrand, Christof Gattringer, Kurt
Langfeld and Uwe-Jens Wiese are gratefully acknowledged. C. Wozar thanks for the
support by the Studienstiftung des deutschen Volkes. This work has been supported by the DFG under 
GRK~1523.
\end{acknowledgments}

\renewcommand{\eprint}[1]{ \href{http://arxiv.org/abs/#1}{[arXiv:#1]}}

\end{document}